\documentclass[%
reprint,
superscriptaddress,
showpacs,
preprintnumbers,
 bibnotes,
 amsmath,amssymb,
 aps,
prl
]{revtex4-1}

\usepackage{graphicx}
\usepackage{dcolumn}
\usepackage{bm}
\usepackage{color}

\begin{document}


\title{Ferromagnetic inter-layer coupling in FeSe$_{1-x}$S$_{x}$ superconductors revealed by inelastic neutron scattering}

\author{Mingwei~Ma}
\email[]{mw\_ma@iphy.ac.cn}

\affiliation{Beijing National Laboratory for Condensed Matter Physics, Institute of Physics, Chinese Academy of Sciences, Beijing 100190, China}
\affiliation{International Center for Quantum Materials, School of Physics, Peking University, Beijing 100871, China}

\author{Philippe~Bourges}
\affiliation{Universit\'{e} Paris-Saclay, CNRS, CEA, Laboratoire L\'{e}on Brillouin, 91191, Gif-sur-Yvette, France}

\author{Yvan~Sidis}
\affiliation{Universit\'{e} Paris-Saclay, CNRS, CEA, Laboratoire L\'{e}on Brillouin, 91191, Gif-sur-Yvette, France}

\author{Jinzhao~Sun}

\thanks{Present address: Clarendon Laboratory, University of Oxford, Parks Road, Oxford OX1 3PU, United Kingdom }
\affiliation{International Center for Quantum Materials, School of Physics, Peking University, Beijing 100871, China}
\author{Guoqing~Wang}
\affiliation{International Center for Quantum Materials, School of Physics, Peking University, Beijing 100871, China}

\author{Kazuki~Iida}
\affiliation{Neutron Science and Technology Centre, Comprehensive Research Organization for Science and Society (CROSS), Tokai, Ibaraki 319-1106, Japan}
\author{Kazuya~Kamazawa}
\affiliation{Neutron Science and Technology Centre, Comprehensive Research Organization for Science and Society (CROSS), Tokai, Ibaraki 319-1106, Japan}

\author{Jitae T.~Park}
\affiliation{Heinz Maier-Leibnitz Zentrum (MLZ), Technische Universit\"{a}t M\"{u}nchen, D-85748 Garching, Germany}
\author{Frederic~Bourdarot}
\affiliation{Universit\'{e} Grenoble Alpes, CEA, IRIG, MEM, MDN, 38000 Grenoble, France}
\author{Zhian~Ren}
\affiliation{Beijing National Laboratory for Condensed Matter Physics, Institute of Physics, Chinese Academy of Sciences, Beijing 100190, China}
\author{Yuan~Li}
\email[]{yuan.li@pku.edu.cn}
\affiliation{International Center for Quantum Materials, School of Physics, Peking University, Beijing 100871, China}
\affiliation{Collaborative Innovation Center of Quantum Matter, Beijing 100871, China}

\begin{abstract}
FeSe$_{1-x}$S$_{x}$ superconductors are commonly considered layered van der Waals materials with negligible inter-layer coupling. Here, using inelastic neutron scattering to study spin excitations in single-crystal samples, we reveal that the magnetic coupling between adjacent Fe layers is not only significant, as it affects excitations up to \textcolor{black}{15} meV, but also ferromagnetic in nature, making the system different from most unconventional superconductors including iron pnictides. Our observation provides a new standpoint to understand the absence of magnetic order in FeSe$_{1-x}$S$_{x}$. Since intercalating between the Fe layers is known to enhance superconductivity and suppress the inter-layer coupling, superconductivity appears to be a more robust phenomenon in the two-dimensional limit than antiferromagnetic order.

\end{abstract}


\maketitle
{\centering\section*{$\mathbf{I.}$ $\mathbf{Introduction}$}}
A common thread for understanding unconventional superconductivity is magnetic interactions \cite{RevModPhysScalapino,TRANQUADA2014148,RevModPhysDai}, about which inelastic neutron scattering (INS) measurements of spin excitations can provide useful information. Since unconventional superconductors are mostly layered materials, a distinct aspect of such information is the nature and strength of inter-layer magnetic coupling, as manifested by magnetic excitations' momentum dependence along the $\mathbf{c^*}$ direction.\\

Taking the well-known spin resonant mode \cite{Eschrig2006} as example, sinusoidal modulation of its intensity along $\mathbf{c^*}$ in Co/Ni-doped BaFe$_{2}$As$_{2}$ \cite{PhysRevLett.102.107006,PhysRevB.79.174527,PhysRevB.81.140510},  NaFe$_{0.985}$Co$_{0.015}$As \cite{PhysRevLett.111.207002}, and ${\mathrm{CeCoIn}}_{5}$  \cite{PhysRevLett.100.087001}, as well as its dispersion in BaFe$_{2}$(As$_{1-x}$P$_{x}$)$_2$ \cite{PhysRevLett.111.167002}, can be attributed to antiferromagnetic spin correlations or interactions between adjacent quintessential layers of atoms that are responsible for both the superconductivity and the magnetism. Such antiferromagnetic coupling is also found in superconductors with two quintessential layers (\textit{i.e.}, FeAs or Cu$_2$O layers) within the primitive cell, such as CaKFe$_4$As$_4$, YBa$_2$Cu$_3$O$_{6+\delta}$ and Bi$_2$Sr$_2$CaCu$_2$O$_{8+\delta}$ \cite{FongPhysRevB.54.6708,TranquadaPhysRevB.46.5561,ChouPhysRevB.43.5554,PailhPhysRevLett.91.237002,XiePhysRevLett,KeimerPhysRevB.75.060502,PhysRevLett.93.167001}. When the inter-layer coupling is negligible, the spin excitations are observed to be nearly independent of momentum transfer along $\mathbf{c^*}$, as found in FeSe$_{0.4}$Te$_{0.6}$ \cite{PhysRevLettQiu}, Rb$_x$Fe$_{2-y}$Se$_2$ \cite{PhysRevB.85.140511}, LaOFeAs \cite{PhysRevB.87.140509},  LiFeAs \cite{PhysRevLett.108.117001}, and Li$_{0.8}$Fe$_{0.2}$ODFeSe \cite{PanNatCommun2017,PhysRevB.95.100504}.\\

\begin{figure}[h]
\includegraphics[width=3.4in]{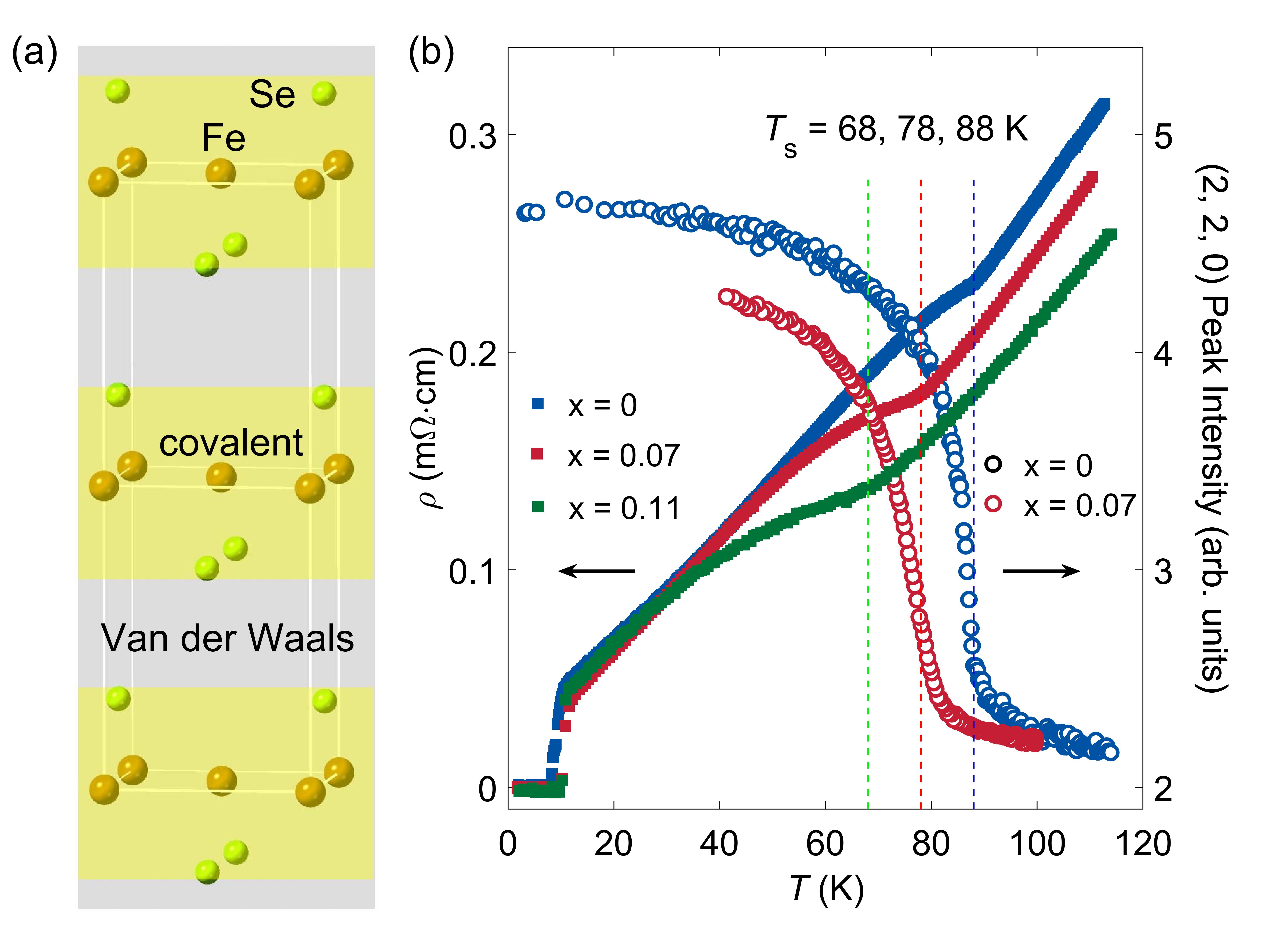}
\caption{\label{Fig1}
(a) Crystal structure of FeSe. (b) Temperature dependence of resistivity and intensity of (2, 2, 0) Bragg peak of FeSe$_{1-x}$S$_{x}$. The elastic neutron scattering for measuring (2, 2, 0) Bragg peak were carried on 4F1 spectrometer for FeSe at the Laboratoire L\'{e}on Brillouin, Saclay, France and the IN22 spectrometer for FeSe$_{0.93}$S$_{0.07}$ at the Institut Laue-Langevin, Grenoble, France. They were both mounted in ($H$, $K$, 0) scattering plane.
}
\label{Fig1}
\end{figure}

FeSe is regarded as two-dimensional material as it is composed of a stack of strongly bonded edge-sharing FeSe$_\mathrm{4}$-tetrahedra layers as shown in Fig.~\ref{Fig1}(a) \cite{Hsu2008}. The weak van der Waals bonding between the FeSe layers allows the material to easily cleave, and it shows potential for use in heterostructure devices with a tunable superconducting transition temperature  $T_\mathrm{c}$ above 40 K \cite{PhysRevLett.116.077002,PhysRevB.95.020503,PhysRevLett.121.207003}. Additionally, the FeSe layers can be intercalated with different charged or neutral spacer layers, or grown as monolayers on SrTiO$_3$, leading to a significant increase of $T_\mathrm{c}$  \cite{LuNatureMater2015,Krzton-Maziopa_2012,GuoPhysRevB2010,GuoNatureCommun2014,YingSciRep2012,WangCPL2012,HeSLNatureMater2013,GeNatureMater2015}.  \\

A unique feature of FeSe is the spontaneous fourfold symmetry breaking observed in the bulk material at the nematic temperature  $T_\mathrm{s}$ = 88 K, driven by electronic or magnetic instabilities. This nematic state, unlike in Fe-pnictides, does not exhibit long-range magnetic order under ambient pressure \cite{McQueen}. The absence of magnetic order in FeSe has sparked interest from various perspectives, including orbital physics \cite{BaekNatMater2016} and quantum magnetism  such as the spin-fluctuation induced spin-quadrupole order \cite{YuRongPhysRevLett2015}, the nematic quantum paramagnetic phase \cite{WangFaNatPhys2015}, the ferro-orbital order in the nematic phase \cite{GlasbrenerNatPhys2015}, the near degeneracy between magnetic fluctuations and fluctuations in the charge-current density-wave channel \cite{PhysRevB.91.201105} as well as the vertex correction \cite{PhysRevX.6.021032}.\\

However, there are few discussions of the absence of magnetic order in FeSe from the viewpoint of the inter-layer coupling. Here we have a try to study its inter-layer coupling revealed by INS and find that the spin resonance mode is highly modulated along $c$-axis momentum transfer with maximum at integer $L$  = 0, $\pm$1, $\pm$2 in an opposite fashion compared to the iron pnictides. Our results suggest a ferromagnetic inter-layer coupling in FeSe.\\

\begin{figure}[h]
\includegraphics[width=3.4in]{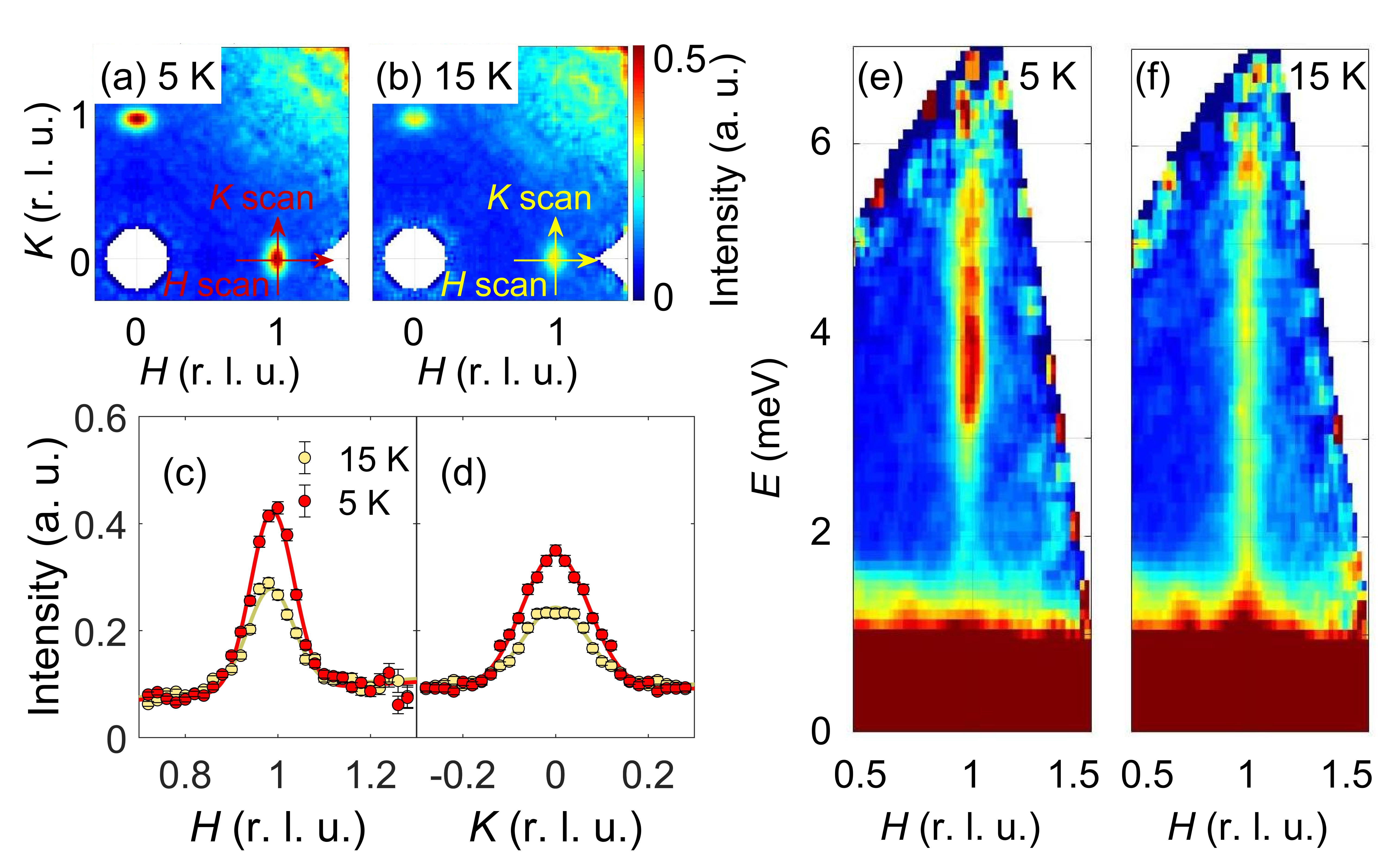}
\caption{\label{Fig2}
(a-b) Constant-energy maps of FeSe$_{0.93}$S$_{0.07}$ obtained at $T$ = 5 K and 15 K in ($H$, $K$) momentum plane with $|L|$ $\leq$ 0.2 and $E$ = 4$\pm$1 meV. (c) Longitudinal scan ($H$ scan) at $T$ =  5 K and 15 K with $E$ = 4$\pm$1 meV, $|K|$ $\leq$ 0.1 and $|L|$ $\leq$ 0.2. (d) Transverse scan ($K$ scan) at $T$ = 5 K and 15 K with $E$ = 4$\pm$1 meV, $|H-1|$ $\leq$ 0.1 and $|L|$ $\leq$ 0.2. (e-f) Energy dependence of two-dimensional slices along $\mathbf{Q}$ = ($H$, 0) direction at $T$ = 5 K and 15 K with $|K|$ $\leq$ 0.1 and $|L|$ $\leq$ 0.2 for FeSe$_{0.93}$S$_{0.07}$. These data are obtained with incident energy $E_i$ = 13.6 meV on a time of flight spectrometer 4SEASONS. Solid lines in (c-d) are Gaussian functions obtained by fitting the respective data. 
}
\label{Fig2}
\end{figure}

{\centering\section*{$\mathbf{II.}$ $\mathbf{Experimental}$ $\mathbf{methods}$}}

The FeSe$_{1-x}$S$_{x}$ (x = 0, 0.07 and 0.11) single crystals have been grown by chemical vapor transport technique \cite{Li_2020} and co-aligned with a mosaic within 5$^{\circ}$ for all the sample arrays as shown in Fig.$\hspace{0.3em}$S1 [See Supplementary Materials]. As displayed in Fig.~\ref{Fig1}(b) S doping causes a slight increase of $T_\mathrm{c}$ and a substantial decrease of $T_\mathrm{s}$ measured by resistivity and (2, 2, 0) Bragg scattering due to neutron extinction effect \cite{Hamiltona02115}, and a full suppression is achieved for x $\ge$ 0.18  \cite{Coldea}. Here and throughout this article, the wave vector  $\mathbf{Q}$ is expressed in reciprocal lattice units (r.$\hspace{0.3em}$l.$\hspace{0.3em}$u.) as ($H$, $K$, $L$). Using the orthorhombic notation of the 4-Fe Brillouin zone, the wave vector, expressed in inverse Angstrom, is $\mathbf{Q}$ = [$(2\pi/a)H$, $(2\pi/a)K$, $(2\pi/c)L$] with lattice parameters  $a$ $\approx$ $b$ $\approx$ 5.32 $\mathrm{\AA}$ and $c$ $\approx$ 5.52 $\mathrm{\AA}$.\\

The triple-axis INS experiments with $k$$_\mathrm{f}$ = 2.662 $\mathrm{\AA}$$^{-1}$ using a focusing pyrolytic graphite (PG) monochromator and analyzer were carried out on the PUMA spectrometer at the Heinz Maier-Leibnitz Zentrum (MLZ), Technische Universit\"{a}t M\"{u}nchen, Garching, Germany, the 2T spectrometer at the Laboratoire L\'{e}on Brillouin, Saclay, France and the IN22 spectrometer using the CryoPAD system at the Institut Laue-Langevin, Grenoble, France. Additional PG filters were placed between the sample and the analyzer to eliminate higher-order contaminations. FeSe$_{1-x}$S$_{x}$ (x = 0, 0.07 and 0.11) single crystals were all mounted in the ($H$, 0, $L$) scattering plane. \\

The time of flight experiments of FeSe$_{0.93}$S$_{0.07}$ were performed on the 4SEASONS spectrometer \cite{KajimotoJPSJ} at the Material and Life Science Experimental Facility (MLF), Japan Proton Accelerator Research Complex (J-PARC), Japan. The 4SEASONS spectrometer has a multiple-$E_i$ capability \cite{NakamuraJPSJ,InamuraJPSJ} with $E_i$ = 10, 13.6, 19.4, 30.1, 52.1 and 116 meV, such that neutron scattering events with a series of different incident energies are recorded simultaneously. During the measurements at $T$ = 5 K, 15 K and 90 K, the sample was rotated about the vertical direction over a range $\pm$55$^\circ$ and a step 0.5$^\circ$ steps. Data accumulated at different angles were combined into a four-dimensional dataset, in which we used the Horace software packages for reduction and analysis. After a careful alignment of the measured dataset with the crystallographic coordinate system using all available nuclear Bragg reflections, the entire dataset was downfolded into a minimal, physically independent sector of the three-dimensional momentum space using the point-group symmetry of the system.\\

{\centering\section*{$\mathbf{III.}$ $\mathbf{Results}$ $\mathbf{and}$ $\mathbf{discussion}$}}

Figure~\ref{Fig2}(a-b) shows the constant-energy maps of spin excitations for FeSe$_{0.93}$S$_{0.07}$ in the superconducting state ($T$ = 5 K) and nematic state ($T$ = 15 K) at $E$ = 4 meV in the ($H$, $K$) plane where the spin excitations are symmetrically located at $\mathbf{Q}$ = ($\pm$1, 0) and (0, $\pm$1) with an elliptic distribution elongated in the transverse direction. The anisotropic distribution can also be displayed  by transverse scan ($K$ scan) and longitudinal scan ($H$ scan) around  $\mathbf{Q}$ = (1, 0) in Fig.~\ref{Fig2}(c-d) as indicated by the arrow in Fig.~\ref{Fig2}(a-b). Compared with the nematic state ($T$ = 15 K), in superconducting state an enhancement of the spin excitations around $E$ $\approx$ 4 meV is accompanied by a suppression of magnetic response below $E$ $\approx$ 2.5 meV as shown in Fig.~\ref{Fig2}(e-f), which is a hallmark of the spin resonance mode in agreement with previous INS studies on undoped FeSe \cite{PhysRevBYiqing,QisiWangNatMater2016,QisiWangNatCommun2016,PhysRevXMA,ChenNat2019}. The elongated distribution along transverse direction of spin resonance in ($H$, $K$) plane is consistent with the previous work on Ba(Fe$_{0.963}$Ni$_{0.037}$)$_{2}$As$_{2}$ \cite{PhysRevLett.110.177002}, CaFe$_{0.88}$Co$_{0.12}$AsF \cite{PhysRevB.107.184516}, ${\mathrm{BaFe}}_{2}{({\mathrm{As}}_{0.7}{\mathrm{P}}_{0.3})}_{2}$ \cite{PhysRevB.94.094504} where the resonance is found to peak sharply at $\mathbf{Q}_\mathrm{AF}$ along the longitudinal direction, but broadens along the transverse direction. Other similar constant-energy maps at $E$ = 2$-$6 meV and energy dependence measurements in $E-K$ space at $T$ = 5 K, 15 K and 90 K are displayed in Fig.$\hspace{0.3em}$S2 [See Supplementary Materials]. When entering the tetragonal state ($T$ = 90 K), the magnetic signals at $\mathbf{Q}$ = ($\pm$1, 0) and (0, $\pm$1) become weak and diffusive on a high background (BG), whereas the scattering signals at $\mathbf{Q}$ = ($\pm$1, $\pm$1) are heavily contaminated by (1, 1, 0) phonon as displayed in Fig. S2(m-r).\\
\begin{figure}[h]
\includegraphics[width=3.4in]{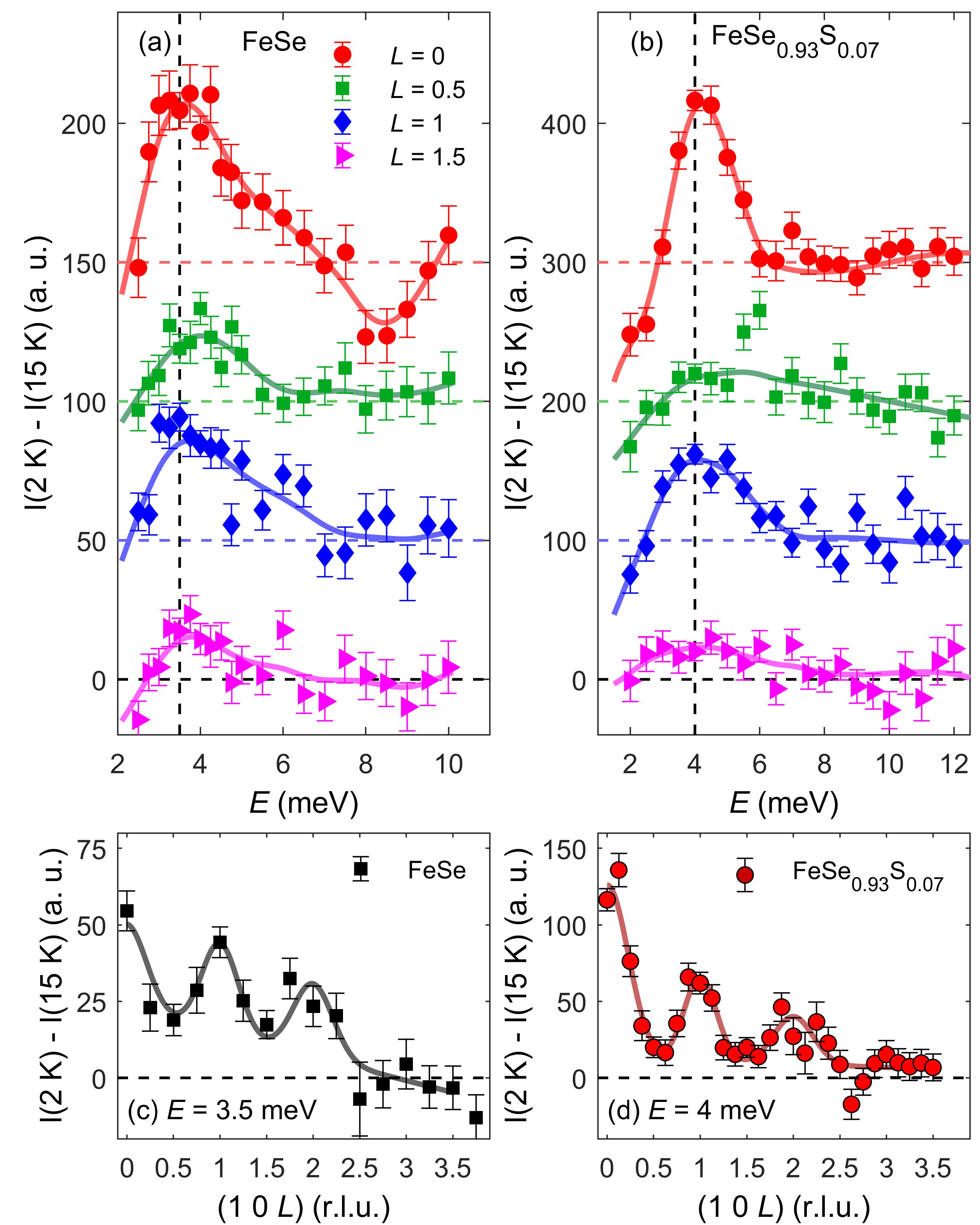}
\caption{\label{Fig3}
(a-b) Temperature difference of spin excitations of FeSe and FeSe$_{0.93}$S$_{0.07}$ between superconducting state ($T$ = 2 K) and nematic state ($T$ = 15 K) at $\mathbf{Q}$ = (1, 0, $L$) with $L$ = 0, 0.5, 1, 1.5 respectively. (c-d) Tempearture difference of spin excitations of FeSe and FeSe$_{0.93}$S$_{0.07}$ between superconducting state ($T$ = 2 K) and nematic state ($T$ = 15 K) along $L$ direction at $E$ = 3.5 meV and 4 meV respectively. Solid lines are the guides to the eyes in (a-b) and multiple Gaussian functions fitted by the respective data in (c-d). 
}
\label{Fig3}
\end{figure}

\begin{figure}
\includegraphics[width=3.4in]{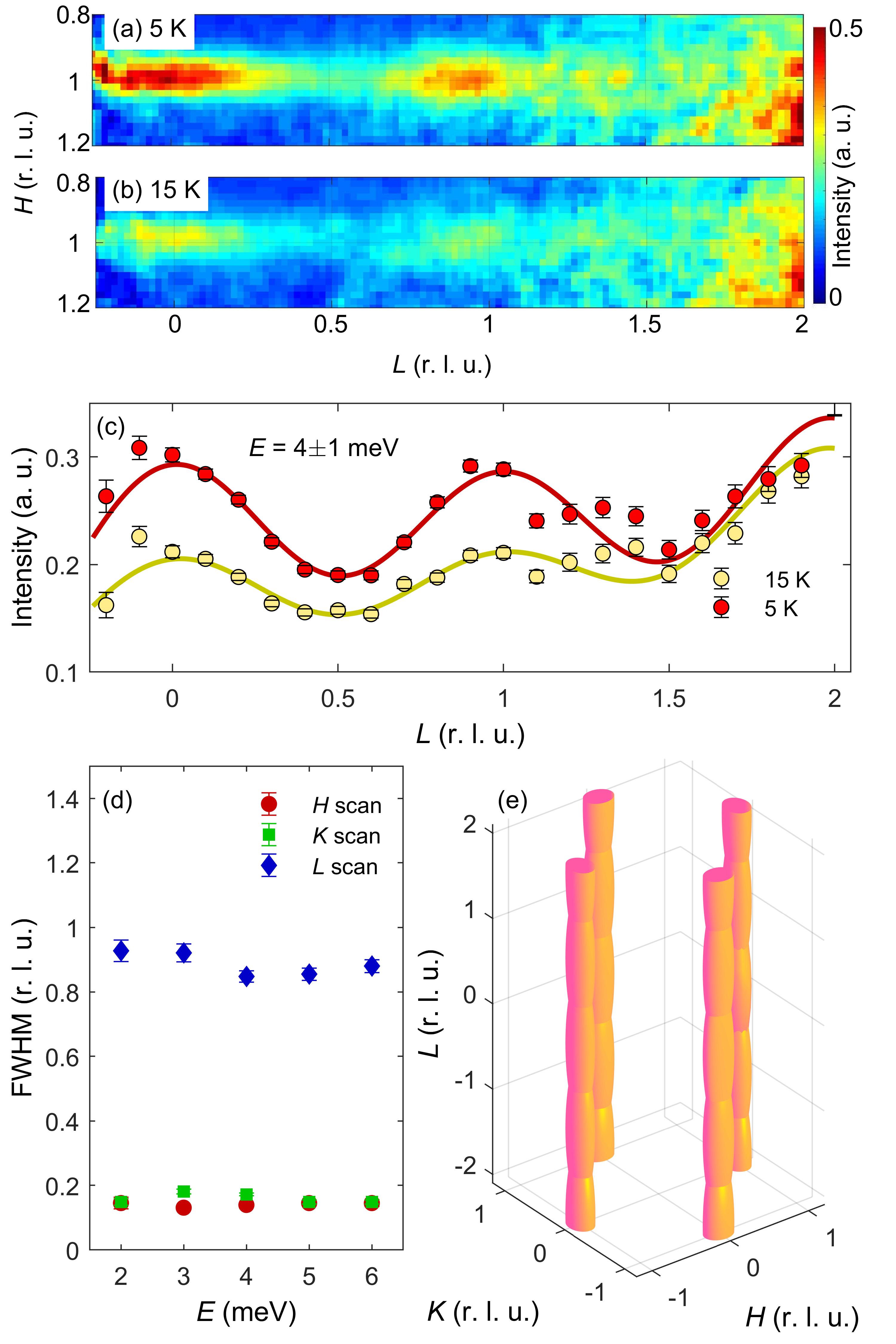}
\caption{\label{Fig4}
(a-b) Constant-energy maps of FeSe$_{0.93}$S$_{0.07}$ obtained at $T$ = 5 K and 15 K and  in ($H$, $L$) momentum plane with $E$ = 4$\pm$1 meV and $|K|$ $\leq$0.1. (c) $L$ scans at $T$ = 5 K and 15 K with $E$ = 4$\pm$1 meV,  $|K|$ $\leq$0.1 and $|H-1|$ $\leq$0.1. (d) Energy dependence of FWHM for $H$, $K$ and $L$ scans fitted by gaussian functions [See Fig.$\hspace{0.3em}$S3 in Supplementary Materials]. (e) The 3D distribution of spin excitation in momentum space using FWHM data from $E$ = 4$\pm$1 meV. Solid lines in are Gaussian functions obtained by fitting the respective data.
}
\label{Fig4}
\end{figure}

After illustrating the anisotropic distribution of spin resonance in ($H$, $K$) plane, we now turn to the $L$-modulation of spin resonance. Figure~\ref{Fig3}(a-b) shows the temperature difference of spin excitations of undoped FeSe and FeSe$_{0.93}$S$_{0.07}$ between superconducting state ($T$ = 2 K) and nematic state ($T$ = 15 K) at $\mathbf{Q}$ = (1, 0, $L$) ($L$ = 0, 0.5, 1, 1.5). The spin resonance mode appears at $E_r$ $\approx$ 3.5 meV in FeSe, slightly enhanced at $E_r$ $\approx$ 4 meV in FeSe$_{0.93}$S$_{0.07}$. Notably, the spin resonance mode presents a stronger intensity at integer $L$ = 0 and 1 than that at half-integer $L$. This $L$-modulation of spin resonance can be further elucidated by the momentum scans along $\mathbf{Q}$ = (1, 0, $L$) at $E$ = 3.5 meV for undoped FeSe [ Fig.~\ref{Fig3}(c)] and $E$ = 4 meV for FeSe$_{0.93}$S$_{0.07}$ [ Fig.~\ref{Fig3}(d)]. In addition, for a higher S doped sample FeSe$_{0.89}$S$_{0.11}$, the spin excitations at $E$ = 4 meV in superconducting state are shown in Fig.$\hspace{0.3em}$S5(e) [See Supplementary Materials].\\ 

This strongly $L$-modulated spin resonance mode in superconducting state is closely related to the anisotropic distribution of low-energy spin excitations in momentum space in the nematic state ($T$ = 15 K). Except for an intensity enhancement due to spin resonance, the anisotropic distribution of spin excitations in momentum space in the nematic state mimics that in superconducting state as displayed in the constant energy maps in ($H$, $K$) plane [Fig.~\ref{Fig2}(a-b)] and ($H$, $L$) plane [Fig.~\ref{Fig4}(a-b)] as well as the $L$ scans in Fig.~\ref{Fig4}(c). The spin excitations reveal elliptic distribution elongated in the transverse direction and a more elongated ellipse in the $L$ direction both in superconducting and nematic state. In order to quantatively clarify the elliptic distribution of spin excitations in three-dimensional momentum space, momentum scans along $H$, $K$ and $L$ directions are presented in  Fig.~\ref{Fig2}(c-d),  Fig.~\ref{Fig4}(c) and Fig.$\hspace{0.3em}$S3 [See Supplementary Materials] respectively. The intensity of spin excitations in the nematic state ($T$ = 15 K) along the longitudinal ($H$), transverse ($K$) and $L$ directions can be well fitted by Gaussian funtions, revealing the  full widths at half maximum (FWHM) value  $\kappa_H$, $\kappa_K$ and $\kappa_L$ as shown in  Fig.~\ref{Fig4}(d).  Comparing the FWHM along $H$ and $K$ directions,  $\kappa_H$ and  $\kappa_K$, respectively, one observes that the latter is slightly larger than the former. With a 6 times larger value  $\kappa_L$ than $\kappa_K$, the $L$ dependence of intensity suggests a much more broadened distribution of spin excitations along $L$ direction, weakened with increasing energy and persisting up to $E$ $\approx$ 15 meV as shown in Fig.$\hspace{0.3em}$S3-S4 [See Supplementary Materials]. According to the FWHM we draw the schematic diagram of three dimensional distribution of spin excitations in momentum space at $T$ = 15 K and  $E$ = 4 meV as displayed in Fig.~\ref{Fig4}(e). The ellipsoids are periodically located at $\mathbf{Q}$ = ($\pm$1, 0, $L$) or (0, $\pm$1, $L$) in a twinned sample where $L$ is integer.\\

We now discuss the implications of our observed INS results. The feature of integer-$L$ intensity enhancement of spin excitations in FeSe$_{1-x}$S$_{x}$  still survives in the nematic state persisting up to $E$ $\approx$ 15 meV and tells us that the spins in nearest-neighbor FeSe$_4$ layers are parallel and fluctuate in phase, which is opposite to the antiferromagnetic order in the parent compound of Fe-pnictides with antiparallel spins of the nearest-neighbor layers along $c$-axis. A ferromagnetic-like inter-layer coupling $\mathbf{J}_\mathbf{\bot}$ has to be responsible for the integer-$L$ intensity modulation in FeSe$_{1-x}$S$_{x}$. \\

The microscopic origin of the absence of the long-range stripe magnetic order in FeSe remains elusive. One scenario is that the stripe magnetic order in FeSe is absent due to the development of other competing instabilities, considering mainly on the in-plane frustrated magnetic interactions \cite{YuRongPhysRevLett2015,WangFaNatPhys2015,GlasbrenerNatPhys2015,PhysRevB.91.201105,PhysRevX.6.021032}. From another point of view, a realistic three-dimensional antiferromagnetic order requires both the spin correlations in and out of plane. The FWHM of spin excitations obtained by Gaussian fitting along $H$, $K$ and $L$ directions is associated with the spin-spin correlation length  $\xi_a$,  $\xi_b$,  $\xi_c$. The FWHM along $L$ scans is 6 times larger than along the transverse scans as the smaller correlation length $\xi_c$ is able to capture the reciprocal space broadening along the long axis $L$. As it is well established that intercalating between the Fe layers can strengthen superconductivity and reduce inter-layer coupling \cite{PhysRevLettQiu,PhysRevB.85.140511,PanNatCommun2017,PhysRevB.95.100504,PhysRevB.87.140509,PhysRevLett.108.117001}, it seems that superconductivity is a more robust phenomenon in the two-dimensional limit compared to antiferromagnetic order.\\

{\centering\section*{$\mathbf{IV.}$ $\mathbf{Conclusion}$}}
In conclusion, the anisotropic momentum distribution of spin resonance in FeSe$_{1-x}$S$_{x}$ mimics the low energy spin excitations in its nematic phase where superconductivity arises, which implies the close relationship between superconductivity and magnetism. The strongly $L$-modulated spin resonance mode inherits the low-energy magnetic scattering in nematic state with maximum intensity at integer $L$. FeSe does not order antiferromagnetically, which is unusual among Fe-based materials. The ferromagnetic nature of the inter-layer coupling might shed light on the absence of magnetic order in FeSe.\\

\begin{acknowledgments}
{\centering\section*{$\mathbf{Acknowledgement}$}}
The work was supported by the National Key Research and Development of China (Grant No. 2018YFA0704200, 2022YFA1602800), the National Natural Science Foundation of China (Grant No. 12004418) and the Strategic Priority Research Program of Chinese Academy of Sciences (Grant No. XDB25000000). One of the neutron scattering experiments was performed at the MLF, J-PARC, Japan, under a user program (No. 2018A0019).
\end{acknowledgments}

\preprint{Preprint}


%

\end{document}